# Protecting sensitive research data and meeting researchers needs: Duke University's Protected Network

Mark R. DeLong, Andy Ingham, Robert Carter, Rachel Franke,

Michael Wehrle, Richard Biever, Charles Kneifel *

9 October 2017

**Abstract**

Research use of sensitive information – personally identifiable information (PII), protected health information (PHI), commercial or proprietary data, and the like – is increasing as researchers' skill with "big data" matures. Duke University's Protected Network is an environment with technical controls in place that provide research groups with essential pieces of security measures needed for studies using sensitive information. The environment uses virtualization and authorization groups extensively to isolate data, provide elasticity of resources, and flexibly meet a range of computational requirements within tightly controlled network boundaries. Since its beginning in 2011, the environment has supported about 200 research projects and groups and has served as a foundation for specialized and protected IT infrastructures in the social sciences, population studies, and medical research. This article lays out key features of the development of the Protected Network and outlines the IT infrastructure design and organizational features that Duke has used in establishing this resource for researchers. It consists of four sections: 1. Context, 2. Infrastructure, 3. Authentication and identity management, and 4. The infrastructure as a "platform."

* Mark R. DeLong is Director of Research Computing, Office of Information Technology, Duke University, 334 Blackwell Street, Durham, NC 27701 (email: mark.delong@duke.edu); Andy Ingham is Senior IT Analyst, Office of Information Technology, Duke University, 334 Blackwell Street, Durham, NC, 27701 (email: andy.ingham@duke.edu); Robert Carter is IT Consultant, Identity Management, Office of Information Technology, Duke University, 334 Blackweel Street, Durham, NC 27701 (email: robert.carter@duke.edu); Rachel Franke is Associate Director, Research Data Security, Social Science Research Institute, Duke University, Campus Box 90989, Durham NC 27708 (email rachel.franke@duke.edu); Michael Wehrle is Analyst Programmer II, Social Science Research Institute, Duke University, Campus Box 90989, Durham NC 27708 (email: michael.wehrle@duke.edu); Richard Biever is Chief Information Security Officer, Duke University, 334 Blackwell Street, Durham, NC 27701 (richard.biever@duke.edu); Charley Kneifel is Senior Technical Director, Office of Information Technology, Duke University, 334 Blackwell Street, Durham, NC 27701 (email: charley.kneifel@duke.edu). Research and products reported in this article were supported by NSF grants ACI-1246042, CNS-1243315, ACI-1440588, and ACI-1443014.

# 1. Context.

*1.1. Increasing need for well controlled and closely monitored computing environments*

Trends in data-rich scholarship, the ubiquity of useful data, and heightened risk of disclosure both to individuals and to institutions have combined to make data protection more of a focus for shared, centralized IT infrastructure and less a decentralized "problem" for researchers and departments to handle independently. The sheer volume of data generated and stored presents its own data management challenges. In a 2014 report sponsored by storage provider EMC, IDC estimated worldwide "digital universe" data sizes would double every two years and amount to around 44 zettabytes (44 trillion gigabytes) in 2020. Of the data assessed in 2013 by IDC, about 43% required some measure of protection because of sensitivity (*e.g.*, corporate financial information, PII, user account information). Of that sensitive data, only 48% was stored in environments that provided data protection commensurate with the need (IDC 2014). Although data held at academic research institutions accounts for a fraction of the total "digital universe," the challenges they pose are still significant. At Duke, centrally managed data doubles every three years. In 2016, Duke's Office of Information Technology managed a total of eight petabytes of data storage, about a quarter of which was used in grant-supported and intramural research; and it should be noted that these data do not include clinical and medical data managed by the Duke Health System.

While universities share characteristics of the great "digital universe," their institutional and network circumstances stand in contrast to the more tightly controlled networks of commercial and business entities – which are of course the IDC report's primary focus. University networks have traditionally been more open and accessible, in part because of history and in part reflecting universities' commitment to "open science" and freedom of academic exchange. That stance complicates implementation of data protection measures, many of which amount to access restrictions placed on data. And yet, campus implementations also must take into account the tradition of free exchange of scholarly information that is a hallmark of university life and academic research.

Tensions arising from the press of increasing data, much of which requires vigilance, and the established mores and practices of academic institutions make up the pragmatic context for the Duke Protected Network. The infrastructure was created to navigate and, as much as possible, resolve these tensions while providing state of the art computational options for campus researchers.

*1.2. Data classification and security*

Data classification often is the starting point for decision-making about data management and, for that matter, about much of the overall data security endeavor, since it frames ways to describe data assets and controls and underlies decisions of access control (see Alberts & Dorofee 2002, Saint-Germaine 2005). Solidifying classification was a prerequisite for designing the Protected Network infrastructure and its features have been derived from risk mitigation and protections that sensitive data require in particular.

The Duke Information Technology Security Office (ITSO) and the Information Security Office (ISO) of the Duke Health System established a three-category data classification standard and defined characteristics of access to data of all types after conducting data risk assessments across campus. The three categories – "Sensitive," "Restricted," and "Public" – correspond to risk levels, with "public" bearing the lowest risk for data approved for release to the general public and "sensitive" being "most restrictive data classification category [...] reserved for data that Duke is either required by law to protect, or which Duke protects to mitigate institutional risk." Under the standard, access to sensitive data is explicitly granted to individuals, while access to "restricted" data is "determined by business process needs" (Duke ITSO 2014) which is, essentially, a role-based framework. The simple three-category approach meets the needs of the Duke enterprise, including research.

Data classified as "sensitive" of course presents researchers and the institution with their biggest management challenges. This classification includes information such as Social Security Numbers, credit card information, PHI, FERPA-protected (*i.e.*, non-directory) data, as well as institutional data of special sensitivity and research data which Duke is contractually bound to protect in specific ways. Controls required for these data – in some cases mandated in government regulations – underlie requirements for technical controls present in the Protected Network, though it is good to note that implemented technical controls do not *in themselves* meet all requirements. The Protected Network standardizes technical controls and allows administrators, staff, and research scholars to build secure processes upon a common foundation of well supported IT infrastructure and administrative process.

*1.3. Real partnership: IT security, IT support, research groups*

The institution determined that organizing security controls around data classification was an efficient approach that would enable Duke to meet regulatory and other third party requirements most effectively, and proceeded down that route when the OIT Protected

Network project was started in 2011. The data classification scheme forms the basis of the infrastructure, and allows the institution to concentrate security efforts on data that present high risk to the institution, thus simplifying compliance with regulations and data use agreements and allowing researchers engaged with sensitive data more freedom to conduct research. The last point – "freedom to conduct research" – is not an insubstantial or incidental outcome, but is baked into the rationale and the execution of the infrastructure. Indeed, without that point, the essential partnership of researchers, IT, and information security staff would be difficult to forge.

Building partnerships in academe isn't for the faint-of-heart, since the ethos of such institutions favors decentralization and researcher autonomy – which make well structured security measures difficult to establish, much less to enforce. Corporate IT models are often seen as too "buttoned down" to be responsive to academic research processes. And yet, as Johnston & Warkentin (2010) observe, "end users are not consistent in their behavioral intentions to comply with recommendations to protect their informational assets. As a result, decentralized IT governance environments, which place a significant portion of decision-making and system management in the hands of the end user, may exhibit a significant[ly] increased vulnerability profile" (see also Warkentin & Willison 2009). That is to say, decentralized IT imparts a higher level of institutional risk, if for no other reason than increasing the variability of IT installations and increasing the likelihood of ineptly implemented security standards. Discovering disclosures in a decentralized environment also is problematic.

The organizational and community framework for the Protected Network is an essential component of the entire security "apparatus" – and one frequently missed by institutions that think of data protection as a technical problem (Biever & DeLong 2015). Not only is initial assent and willingness required from researchers, research administrators, security professionals, and IT staff, but a communication mechanism for ongoing involvements and interactions is necessary. At Duke, research group manager meetings take place on a bi-monthly schedule. These are attended by IT and data security staff from departments, centers, and schools whose researchers use the Protected Network and university leaders in information security and research computing. The group discusses use cases, tool and resource development, changes to systems and processes, and generally consults on matters relating to use of sensitive data in research and ways to engineer scalable and flexible responses to researchers' computing needs for data-sensitive work.

**2. Infrastructure.**

*2.1. Air-gaps and "little boxes" v. structured network traffic and virtualized environments*

Strategies for isolating sensitive data reflect tools that are available at the time when protection of data "becomes a problem," and in many cases technological inertia or regulation have frozen methods of data protection in place. For example, some data agreements still require an "air gap" between computers holding data and other networks in addition to other physical security requirements and modes of data transport. The strategy, of course, can work quite well to protect data, though certainly not perfectly (as a caution, see Falliere, Murchu & Chien 2011; Guri, Monitz & Elovici 2016, among others), by making security measures more physically apparent and usually requiring physical access or proximity to accommodate restrictions. Unfortunately, "air gap" systems also constrain research to methods that can be executed on hardware that happens to be in place. Given the quick progress in technological development, today's amply outfitted hardware quickly becomes tomorrow's woeful "little box." The term "air gap" itself has acquired a bit of a metaphorical hue as well, since some companies claim "that a network or system is sufficiently air-gapped even if it is only separated from other computers or networks by a software firewall" (Zetter 2014).

In part to avoid the inevitable constraint of "little boxes," the design of the Protected Network aggressively uses virtualization of IT resources to make the environment suitable for – and scalable to – the diversity of research projects that are conducted within it. Projects require tailored computational tools depending on the software that is needed, the size of datasets, the computational power that is required to execute analysis, and interfaces or operating systems that are familiar for researchers (or required by software tools). In addition, virtualization allows changes over the course of a project, so that IT infrastructure can align with changing – and often intensifying – computational needs. Thus, this fundamental virtualization strategy for provisioning resources increases the efficiency of research by allowing for dynamic sizing and resizing of resources in accord with computational demands. Indeed, whole computational setups can be tested and, as the case may be, abandoned and reconfigured without regard for hardware limitations and encumbered capital expenses that are part of often sizable equipment purchases.

*2.2. Partitioning of resources in the Protected Network*

The partitioning of basic IT resources is accomplished through use of hypervisors for supplying compute cycles (2.2.1), segmentation of storage (2.2.2), network isolation and segmentation through virtualization techniques and software-defined networking (2.2.3), as well as use of applications that handle user and group authorization appropriately (2.2.4). We have also explored the use of separate cloud infrastructure to see how cloud resources can be integrated into the Protected Network environment, preserving the partitioning while also being tightly knit into the on-premise infrastructure (2.2.5).

2.2.1. <u>Hypervisor for flexible compute.</u> Currently, virtual machines (VM) are created using VMware ESX as the hypervisor, and that hypervisor is extensively used through all of the Duke Research Computing infrastructure, including the high-throughput and high-performance clusters and other on-demand scientific computing services. The homogeneity of the virtualization framework greatly reduces the balkanization of computing resources, making deployment of cluster computing systems (for example) within the Protected Network simpler and allowing for the temporary staging of machines and quick changes of RAM and CPU-core configurations. In effect, Protected Network resources can be scaled because a large and malleable computational resource underlies it, in the manner of "cloud"-like raw materials of compute cycles and storage. Virtual machines have the advantage of easily being destroyed and reconstituted, which adds flexibility and tightens economy of use while also helping to secure data. Destroying a VM in effect destroys data that may have been left in some state on a physical machine.

Systems provisioned in the Protected Network may share a hypervisor with a VM that is outside of the Protected Network, though whole physical machines can also be dedicated for the Protected Network with the effect that VMs are isolated from other, unprotected, networks at the hypervisor host level (see Riddle & Chung [2015] for a survey of hypervisor security in cloud environments). The current infrastructure uses VMware ESX (https://www.vmware.com/products/esxi-and-esx.html) as the hypervisor, though other hypervisors, such as KVM, Xen, or Hyper-V can be used. Services devised at Duke for orchestration and deployment of virtual machines are not tied to any specific hypervisor product.

2.2.2. <u>Segmentation of storage capacity</u>. Although concerns of data disclosure certainly are evident in the configuration and maintenance of VMs, persistent storage of sensitive data on data storage equipment and restrictions on their access figure prominently into the design of the infrastructure. Encryption of data-at-rest is possible as an option, though we do not consider encryption of data-at-rest as a particularly effective tool for data protection on its own.

Storage resources are logically isolated using network controls and access control lists (ACLs). Network attached storage uses only CIFS protocol, and storage devices for the Protected Network in general house data for several logically separated projects. Circumstances of one project required use of iSCSI, and isolation of the data in that case was accomplished by dedicating physical hardware within the Protected Network solely to that project, a circumstance that limited the ability of the project to have storage resources resized according to needs. Storage was exported to the project's virtual machine in the Protected Network. As a rule, NFS is not acceptable in the Protected Network.

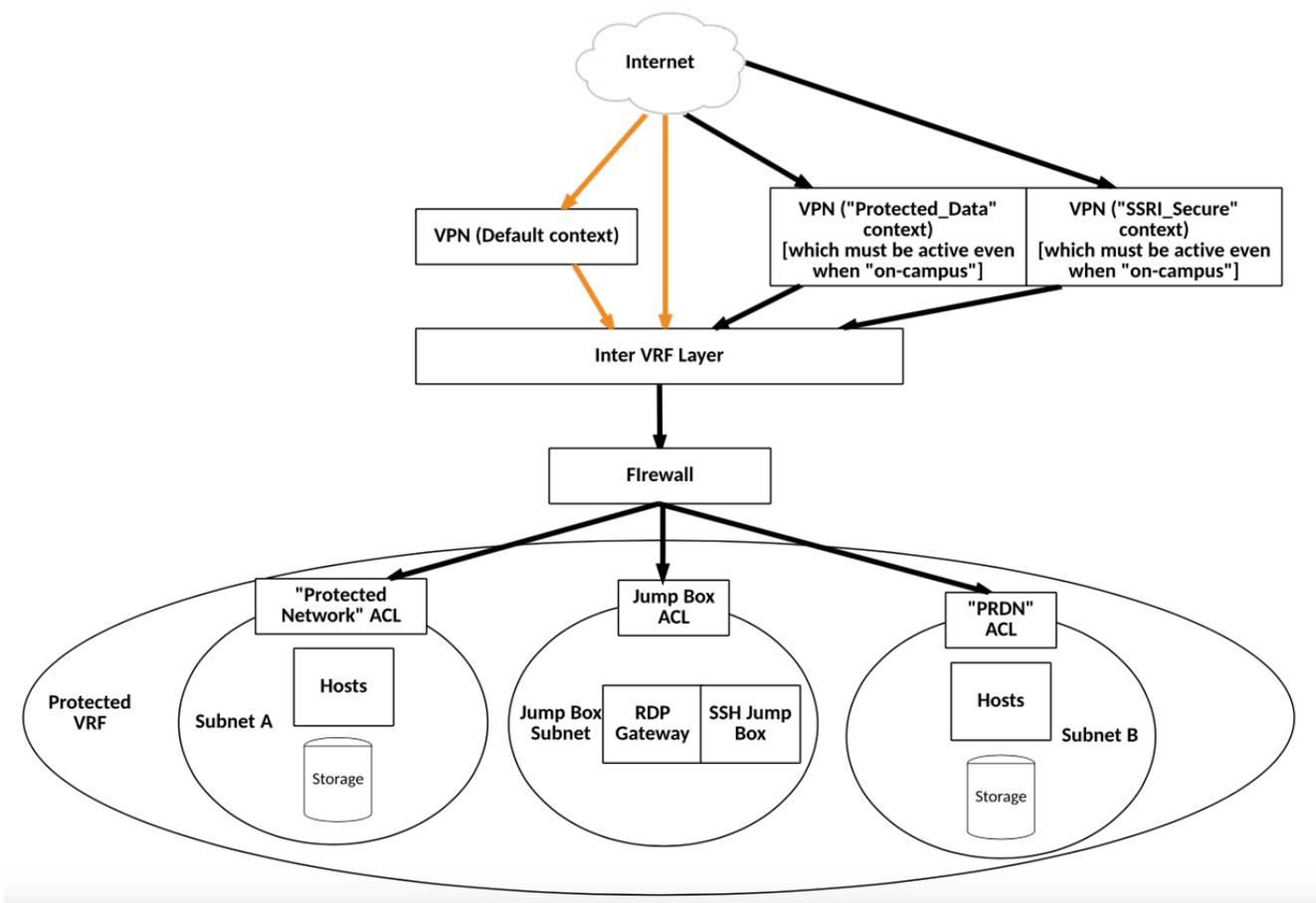

**Figure 1: Virtual Private network (VPN) access.** The Protected Network has two VPN contexts that constrict access to any of the networked machines, with pathways designed for the Protected network displayed with **BLACK** arrows. Access, once achieved, is further narrowed with ACLs that are specific to individual hosts, with access granted to group members. That is, within the enclave, hosts and storage are additionally restricted. Some parts of the Protected Network further segment access by limiting privileges once users gain access; this is done by limiting VPN access and supplying a limited version of Remote Desktop Protocol (RDP).

2.2.3. Network partitioning and host isolation. Data storage and computational cycles are, of course, endpoints that provide some sort of service; the essential element that makes these services consumable is access, which is fashioned by the connections and the rules governing them. This structure of connections is the key element that distinguishes the Protected Network and is critically important to creating an environment appropriate for use of sensitive data.

The Protected Network concentrates access through monitored gateways, currently VPN, RDP, and, in some cases, SSH. User access directly to hosts or storage in the environment is forbidden by firewall rule and network design, and the Protected Network has a designated VPN context separate from all others at the University (Figure 1). The Social Science Research Institute (SSRI) has a separate VPN context that admits users to a subset of the overall Protected Network (described in section 4.1).

Duke's MPLS-enabled networking infrastructure allows for the creation of "virtual routing and forwarding" (VRF). The Protected Network is set up as a VRF designed 1) to limit inbound access to hosts residing in the Protected Network by constraining access from outside through certain closely monitored gateways and 2) to establish a network that functions with minimal network connections, outbound and inbound, with close monitoring for "housekeeping" services that do traverse the VRF boundary.

As with the use of a hypervisor for supplying computational capacity, the use of the VRF technology allows flexibility in network configuration. That flexibility of network deployment means that widely separated machines, physically located remote from other protected machines, can be networked in virtual manner so that they reside "close together" (*vis-à-vis* the network) in a subnet and within the network configuration that the Protected Network VRF enforces. VRF technology underpins some of the innovation of a protected research environment currently being designed by Duke University and Duke Health System (described in section 4.3).

Virtual routing and forwarding features are also represented in Software-Defined Networking (SDN), though with a dramatically different network infrastructure that disassociates network control from the switching equipment, in effect turning network control into a software concern and simplifying centralized network management (Kim & Feamster 2013; Kreutz *et al.* 2015). Duke's SDN installation supports an extensive "Science DMZ" that began with work done on grants from the National Science Foundation. Work most recently done with SDN has moved the Science DMZ to a multi-institutional scope, linking SDN networks and using research and education networks regionally and even internationally, using a research and education network to reach South America.

The Science DMZ seeks to address the problem of data mobility, especially as it relates to Big Data science and high-performance computing installations. However, the use of SDN in the context of sensitive data has to do with the transfer of typically large data over privileged and isolated routes and, conceivably, the incorporation of well managed devices into the Protected Network – as may be desirable for special computing equipment or data-producing devices that typically reside in scientific labs or shared resources.

In three NSF-supported projects (ACI-1246042; CNS-1243315; ACI-1440588), work was done on an interface for secure control of SDN networking in a web-based application called "Switchboard" (McCahill 2015a). Switchboard, as its name implies, is a human interface that connects endpoints via an SDN-controlled route; people who are authorized to control the endpoints use Switchboard to initiate the SDN connection. Switchboard creates the rules inserted into an SDN controller (Ryu; https://osrg.github.io/ryu/) and maintains the state of the ruleset. Switchboard has been developed so that it can serve as a means to span controls over SDN networks in the manner of an SDN exchange (SDX), and an SDN exchange has been experimentally established in the Research Triangle (NC) region.

The ability of SDN to carve out routes that link endpoints and allow for routing that bypasses middleboxes and firewalls that impede traffic flows has obvious benefits for transport of large data, though with the caveat that security protection afforded by firewalls and IPS devices is skirted. Switchboard, however, imparts a measure of protection by requiring particular and specific authorization of use of endpoints by people who are responsible for the endpoints. Encryption of data over the route would, of course, be done.

Another NSF "CICI" grant (ACI-1642140) currently led by Jeffrey Chase (Department of Computer Science) is devising means to increase security of high-speed networks without taxing performance typical of traditional traffic inspections common in "regular" campus networks. The system uses Bro (bro.org; Paxson 1999) and also integrates a system to allow logical analysis of trust for authorization (Thummala & Chase 2015).

2.2.4. Partitioning application "services." The overall framework of the Protected Network can incorporate any information technology or service that can be reliably segmented so that users can consume data and computational resources in a manner that prevents interactions between projects or among users who are not authorized to use a project's resources. This includes applications, such as Gitlab which has been installed within the Protected Network for use by members of different sensitive projects. Users' storage and environments are isolated at the project level. In cases where applications do not have access controls to provide users or groups individually isolated workspaces, applications can be installed for specific groups on a group's individual host. This, of course, complicates management of machines and software but also gives researchers more options for software environments required for their projects.

Systems administration tools that reside outside of the confines of the Protected Network such as patching systems, vetted software repositories, and monitoring systems are available by special firewall exception or by proxy services. Of course, access to outside resources is allowed for certain well documented administrative functions executed by IT staff (*e.g.*, system updates and monitoring). Squid (http://www.squid-cache.org/; Wessels and Claffy 1997) has been set up to provide for some of these, and the University ITSO will consider research project uses of the proxy services under certain circumstances, such as scheduled data downloads from a data provider. Squid allows the ITSO to narrow the source of web-based data downloads quite precisely.

2.2.5. Integration of cloud resources. Although, at present, active projects supported by the Protected Network all consume local infrastructure, we have tested integration of cloud-based infrastructure as a means of expanding services without undermining the protections for the computing environments – long a priority for cloud vendors such as Amazon Web Services (AWS) and Microsoft Azure and scrutinized by the research community (see, for illustration, Amazon Web Services 2017a & 2017b; Ayad, Rodriguez, & Squire 2012; Nellore, *et al*. 2016; Zhang, Cheng, & Boutaba 2010). In 2016, we used Amazon Web Services to build a "virtual private cloud" (VPC) extension of the Duke network, linking AWS to a specific subnet within the Duke Protected Network.  We were able to fully incorporate our centralized authentication (Shibboleth) and authorization (Grouper) mechanisms into the AWS Management Console and to successfully test basic provisioning of S3 and EC2 instances. Additionally, we were able to demonstrate that similar Shibboleth and Grouper integrations work as advertised for command-line (CLI) interactions with specific instances via the published API.  Work still to be done includes (1) investigating AWS services such as

CloudTrail and CloudWatch to allow events notifications from EC2 instances to be incorporated into OIT's current systems information and event management environment, (2) investigating AWS services such as Trusted Advisor and AWS Config to allow for more comprehensive visibility into performance, reliability, cost efficiency, and security compliance for AWS-in-Protected Network VM instances, (3) addressing issues of scalability (including adding a second, mirrored VPC tunnel) since there may be AWS-imposed constraints on Virtual Private Cloud network, or other technical constraints to complicate broad deployments, and (4) mapping out and implementing a strategy for managing a portal (complete with billing management that allows a functional, secure, auditable, and manageable deployment across a broad and diverse campus community) that can also integrate well with automated provisioning mechanisms currently deployed at Duke.

Special services from cloud vendors, such as Amazon's "GovCloud," are designed to meet specific security and regulatory requirements for data and computing. These services make it possible to use cloud services infrastructure to comply, in part, with regulations as stringent as ITAR, FIPS 140-2, FISMA, PCI DSS Level 1, and the like.

**3. Authentication and Identity Management**

While the Protected Network as a whole comprises a group of machines, in effect this grouping is an administrative convenience, since access to individual machines and resources resolves in a more constrained fashion. That is, access within the Protected Network is granted to individual machines and individual data sets relating to a project – in effect further isolating project resources within the Protected Network – while the larger Protected Network provides a perimeter and a single, well monitored point-of-entry into the protected environment. That additional isolation is accomplished by using access control lists (ACLs).

*3.1. Use of centralized authentication but "local" control of access provisioning.*

Access to the environment leverages centralized identities ("NetIDs"), and all access requires multi-factor authentication (MFA) via Duo (Internet2; duo.com) or Yubikey tokens (https://yubico.com). Access and authorization, using InCommon mechanisms offered by Duke's Office of Information Technology, are managed by researchers or by their data security staff, in effect establishing a shared security model with many responsibilities assumed by those "close to the scholarship." Typically, PIs or their designees (defined as "Data Stewards" in the Data Classification Standard) grant or revoke access to data and computational resources, following data use agreements and IRB requirements. These individuals manage access by Grouper group (http://www.internet2.edu/products-services/trust-identity/grouper/) membership of users whose passwords are authenticated via Kerberos within Duke's Active Directory environment.

In the case of an area of the Protected Network managed and maintained by Duke's Social Science Research Institute (SSRI), Grouper is further used to classify users by approved method of access and thereby grant more finely defined privileges for their use of the environment. Guidelines from SSRI data security staff help with compliance and take some administrative burden from researchers (see section 4.1).

*3.2. Accommodation of external collaborators.*

Since research often involves collaborators from institutions other than Duke, access provisioning is done with two methods: use of federated authentication (available by arrangement with other InCommon institutions) or provisioning of a Duke "NetID" as an affiliate, a process that qualifies access through a Duke-based sponsor who bears responsibility for accuracy of the record and maintaining currency of active collaborators as well as complying with appropriate approval processes. With support from the National Science Foundation (ACI-1440588), Duke OIT and campus researchers greatly expanded the scope of web-based federated authentication using the well established Shibboleth tools (Internet2, Carter 2015). Essentially, Duke developers subjected entire operating systems – Windows and Linux – to Shibboleth authentication in web browsers, making any Windows- or Linux-platform application into a tool accessible and usable within a web browser. Previous to this development, web "domesticated" applications were web-based by design (*e.g.*, R-Studio or Wordpress). With the work accomplished during the grant, any Windows- or Linux-based application (*e.g.*, Microsoft Office or Matlab) can be used on a machine accessible by Shibboleth and presented within a web browser.

3.2.1. "Domestication" of operating system platforms and applying web-based federated authentication tools. The federation work done in the NSF-sponsored project in part was an example of "domestication" – or "the process of externalizing authentication, authorization and group management from applications."[1] The project's focus on domesticating whole operating

---

1  A fuller description from SURFnet is useful: "Domestication is described as the process of externalizing authentication, authorization and group management from applications. Domesticated application[s] typically use a[n] external authentication source like for example system platforms, however, dramatically extended the range of software available "on the web" via a simple modern web-browser. Mark McCahill and Robert Carter did the bulk of the work, with McCahill (2015b) focusing on Linux and Carter (2015) on Windows in a system called "Proconsul." An open-source project from the Apache Foundation, called "Guacamole," is currently "incubating" and was considered for adoption for the project (https://guacamole.incubator.apache.org/). McCahill's and Carter's systems essentially "glue" various already available applications and services, such as xVNC, authentication mechanisms (both Duke's centralized systems and those local to machines), Docker (for Linux), and virtual machine orchestration mechanisms. Both the Linux and Windows systems "map" users to virtual machines or to container instances, so there is a level of abstraction separating a user's identity in Duke's IdM system (*i.e.*, the "NetID" of a "Real Persistent User") and the arbitrarily created "user" of a Linux or Windows instance. Because these accounts are unified in a database, users' activities on VMs, of course, can still be traced to individuals.

The entire set up "externalizes" the authentication mechanism from the operating system to the web, allowing for use of Shibboleth, which of course is web-based authentication that also supports multi-factor authentication. Successful "Shib" authentication links the authenticated Real Persistent User to an arbitrary user on the virtual machine. The linkage is recorded in a database, so that sessions can be reinitiated in a preserved state and retained for the record. In order to maintain a continuity of a user's access to network attached storage (such as a mapped "Windows Network Share"), some associations of network storage controlled by the Real Persistent User and the arbitrary user (and that user's short-lived credentials) are also created on-the-fly.

The association of a Real Persistent User and the arbitrary user is, thus, maintained by the system, but with clear distinctions that make the system work seamlessly and also provide security benefit: the Real Persistent User's identity and credentials are never available to the VM; likewise the arbitrary user's "identity" and credentials on a Windows or Linux VM are

---

a SAML based identity federation, and communicate with group management and authorization systems to retrieve additional information on the authenticated user, his/her roles and rights…. Domesticated applications enable single sign-on features for users, as well as the ability to share group context between multiple applications. From the service provider point of view, externalizing identity and group management eases the burden of maintaining these kind[s] of account data." Cited in DeLong, 2016, at reference: https://wiki.surfnet.nl/pages/viewpage.action?pageId=3571713 (Accessed Feb 16, 2016; but on Apr 26, 2017, "not found").

never revealed to the Real Persistent User, either. Moreover, the VM's arbitrary user's credentials (in the Duke implementation at least) are ephemeral, lasting only for a single session before being destroyed.

The distinction of Real Persistent User and the arbitrary user on the virtual machine allows for greater control of access, since normal (and typically long-lived) permissions of the Real Persistent User can be completely changed and shaped by the permission granted to the arbitrary user for a specific session on a VM. In Proconsul, the orchestration mechanism for Windows machines, this distinction of Real and arbitrary users brings both complication and security benefit. The abstraction layer between the authentication systems (web-based Shibboleth and the local Windows OS) complicates mappings of the Real Persistent User's home directory provided by the University, because within the context of the Windows VM the Real Persistent User's identity is re-represented by an ephemeral arbitrary user without the privileges that the Real Persistent User enjoys. That is, from the vantage point of the university's centralized storage system, the Real and the arbitrary users are actually completely different, with the result that file and directory access does not align for them. This disparity of user identity is managed in Proconsul by manipulating group membership; Proconsul adds the arbitrary user of the virtual Windows machine to the same group as the Real Persistent User. Access is thus managed at the group level, where the Real and arbitrary user permissions can be brought into alignment. See Figure 2 for a visual representation of the Proconsul system.

But the arbitrary user can be assigned completely different privileges than the ones granted to the Real Persistent User, and that realignment of privileges can have great security benefit because the privileges are short-lived. This means, in effect, that permissions can be crafted (for example) to permit certain data accesses only via a VM provided through the Proconsul system, narrowing Real Persistent User's options and permissions to a very specific machine set up under very specific and temporary circumstances. A concrete example of the benefit exploits Proconsul's abstraction layer between Real and arbitrary users as well as the mediating abstraction layer separating the web and Windows VM. Duke's Office of Information Technology uses Proconsul's abstractions of authentication and user identities to limit the exposure of privileged user credentials to the Windows OS (which is targeted by Pass-the-Hash exploits) and by doing so limits the "kill-chain" of compromises of Active Directory services (Hummel & Niem 2009; Oberle, *et al*. 2016).

The developer, Robert Carter, called an early version of the Proconsul system "Schaufenstern" for good reason: the system separates the "outside" from the "inside" by a "pane of glass," metaphorically speaking, through which you can see but which also prevents direct interaction. (*Schaufenstern* is German for *display windows* or *shop windows*.) That separation has security benefit, for by separating the mechanisms of authentication and coordinating them by other back-end means, computers can be shielded from exploits like "Pass-the-Hash" which exploit the persistence and transmission of credentials in networked Windows systems (Hummel 2009). Although a Pass-the-Hash attack cannot be entirely prevented, the temporary nature of the Windows credentials in the system severely limits the timeframe for an attack. For all practical purposes, the timeframe is a single user session.

The technology also has some advantages in reducing the risk of connections from "unmanaged" devices – *i.e.*, computers not managed by IT staff to a set of standards or policies, and which are therefore more prone to compromise. The benefit comes from the "pane-of-glass" separation from the target VM located within the Protected Network. Moreover, the Linux or Windows desktops of the virtual machines are presented in a modern web browser, providing the added benefit that no client-side software, like VNC or "Remote Desktop" is required. Restrictions of "copy-n-paste" features are also possible by disabling them in JavaScript. Full use of web-browser-based access to the Protected Network has not yet been implemented.

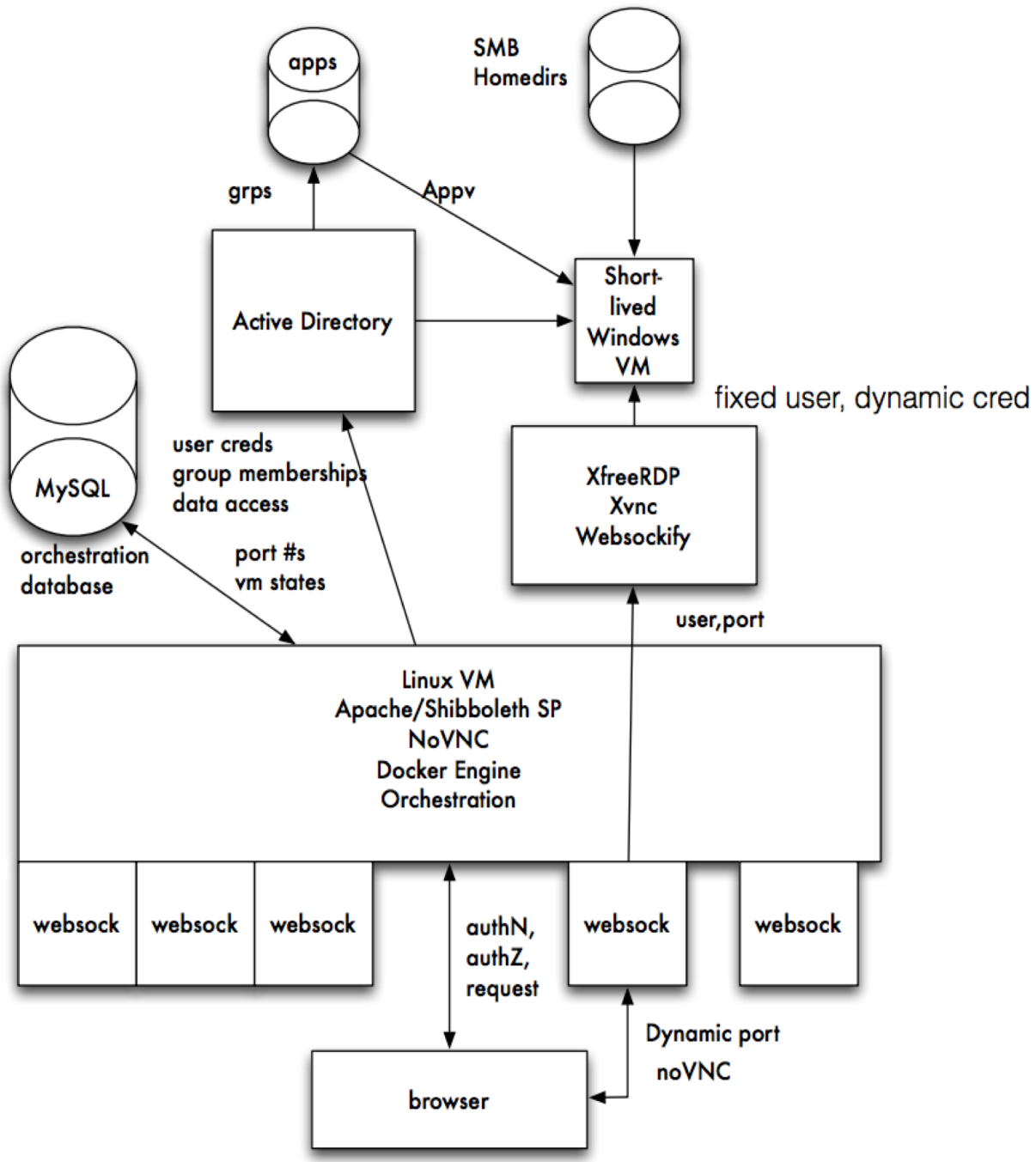

**Figure 2: "Proconsul" infrastructure for web-browser-based Windows.** From a user's perspective, the process of accessing a "domesticated" Windows VM is: 1) A user from University A points her web browser (at the bottom the figure) at a page to gain access to a Windows VM offered by University B. 2) University B determines whether the user qualifies for access by seeing whether University B counts her as OK, either by consulting internal records or rules and asking University A to confirm the user's credentials ("authN, authZ request"). 3) University B maps the user to a temporary username on a virtual machine and passes the desktop of that machine through a system using RDP and VNC to render the desktop to the user in a web browser. The user's claim on the virtual machine will persist, but the credentials for access to the virtual machine is unknown to the user. 4) Storage resources required by the user are made available to the temporary user on the Windows machine. 5) At the close of the session, records of the session are kept, but the temporary username and credentials are destroyed. The virtual machine (now inaccessible with the destroyed credentials) is retained for a certain period, so that the state of the machine persists and can be accessed by the user — though using another set of temporary credentials created for the next session (DeLong 2016; diagram from Carter 2015).

**4. Duke Protected Network as an infrastructure "platform"**

While most research projects using the Protected Network require little more than a secure place for data storage and for computational analysis, some projects and university organizations have used the Protected Network as a platform upon which to build research infrastructures that shape and augment resources to fit specific purposes – in effect extending and creating a unique application of the combined features of the Protected Network. The Protected Network infrastructure makes it possible to develop new services that fit certain research activities that use sensitive information. This feature particularly expresses Duke's over-arching philosophy of devising research IT around research intentions and processes (of course mindful of data classification), rather than the reverse: conforming research workflows and processes to a rigid and monolithic IT infrastructure. In effect, this general stance of research IT at Duke allows for innovative IT infrastructure to become a true research *product* complementing more conventional academic research products, such as publications, data sets, or software.

We present three examples below, one from a major institute at Duke (4.1), another that is designed for a specific research project (4.2), and a third application that uses Protected Network features as a model for a similar environment designed for biomedical/clinical research (4.3).

*4.1. Special application: Protected Research Data Network (PRDN)*

The Protected Network is designed to simplify meeting computation and storage needs of university researchers and to turn many security requirements into technical controls. Meeting administrative requirements of data providers, campus and health system IRBs, information security offices, and the health system's compliance office is often more nuanced and complicated. Mindful of the effort required for these administrative requirements, Duke's Social Sciences Research Institute (SSRI, https://ssri.duke.edu/) supports social sciences researchers from a number of Duke academic departments and has created the Protected Research Data Network (PRDN) which is an isolated segment of the Duke Protected Network. The PRDN embodies a well balanced protected computing environment, since it uses technical controls available in the Protected Network, calls upon the talents of SSRI's IT personnel, and links these with data security and compliance talents at the institute and the broader University in order to strengthen administrative controls for the environment to meet data providers' requirements. In short, the PRDN emerges from a vital partnership of managed information technology, data security, and researchers. The SSRI PRDN was a logical extension of the development of the Protected Network, since it would have been inefficient, and perhaps impossible, to require all researchers on campus to use a specific research environment, without regard for the data they use and their methods and habits of analytical tools. The focus on social sciences research simplified early deployment and support of research, if for no other reason than the research activities bear a certain familial similarity. Currently, the diversity of researchers using the PRDN as a protected research data enclave has expanded beyond the social sciences to engineering, bioengineering, medical departments and projects involving export controlled data, among others. The SSRI PRDN team has managed data sets from over 75 different data providers, in support of 125+ different projects and 200+ individual researchers in the past three years. Administrative and technical controls implemented in the PRDN meet the requirements of most data providers so that little control customization is necessary from project to project. The technical and administrative controls are regularly evaluated against current requirements to ensure that the PRDN meets the majority of researchers' protected data needs. Computation scales to very large RAM VMs and GPU VMs to support machine learning projects; PRDN storage capacity in 2017 has neared 100TB.

The PRDN exploits some distinctions in the overall Protected Network environment in order to differentiate groups using the PRDN and supply them with different privileges over data sets with which they work. The PRDN has its own VPN context which is closely monitored by PRDN staff. Groups can be differentiated by which mode(s) of access they use. While members of research groups can only use data to which they are explicitly granted access, actions they can take with data depend on whether they are authorized to use the VPN or not. If they are not authorized for VPN usage, they are limited to using Remote Desktop Protocol (RDP) provided through a central Terminal Services "jumpbox." RDP users cannot remove data from the PRDN, and "copy-and-paste" features in the environment have been deactivated. Data stewards – usually the data managers or faculty PIs – can use the VPN context, using an endpoint that is managed by Duke IT staff to meet university information security standards for sensitive data. Thus, by exploiting the distinction between VPN and RDP technologies, the PRDN has created a "technical control" on data access enforced by careful administrative control over group memberships – that is, over who is permitted to use VPN and who uses RDP.

In addition to the creative implementation of VPN and RDP controls, other standards are implemented as well, such as a consistent set of software on all VMs using Duke-provided central repositories that have been customized for the PRDN. Reports are created for compliance and to alert researchers of resource efficiencies or deficiencies on their respective systems.

*4.2. Special application: Research environment for synthetic data research*

The NSF-sponsored "CIF21 DIBBs: An Integrated System for Public/Private Access to Large-Scale, Confidential Social Science Data" (ACI-1443014) illustrates how the Protected Network could support the development of a specialized system and turn an IT innovation supporting a research project into a model and replicable research product. The project was conceived early on as being housed in the Protected Network, using SSRI's PRDN, primarily because the nature of the data required significant protection and vigilance. Its data is from the US Government Office of Personnel Management (OPM) CPDF-EHRI (Barrientos, *et al.* 2017; Reiter 2017). Personnel from Duke's OIT and SSRI were engaged very early as the PIs began framing the grant proposal. Since the award in 2014, IT staff have been directly engaged with the execution of the project. This close engagement of information technology staff with the research team has enabled an expansion and tailored provisioning of resources in the PRDN so that resources directly support the DIBBs research workflows.

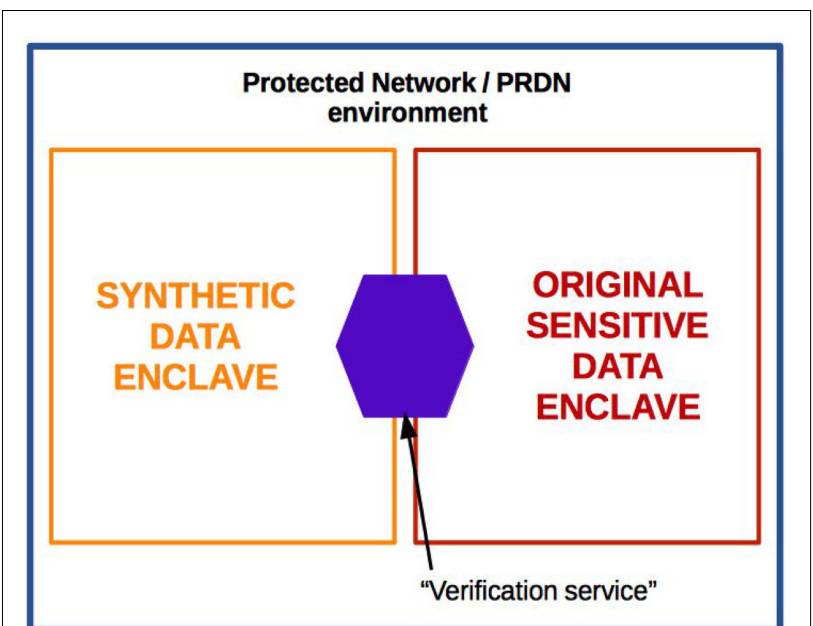

Figure 3: **The integrated system.** In order to provide a protected area for development of the system, all components are encapsulated in the Duke Protected Network / PRDN (**BLUE** box). In other environments where synthetic data is able to be made public without disclosure concerns, such a complete isolation would not be required. The original sensitive data is protected (**RED** box), and those data would always be tightly controlled. In the case of the DIBBs grant, this network area contained the OPM data and was not accessible by individuals who were granted access only to the synthetic data. Synthetic data modeled from the original data (**ORANGE** box) allows access in a less restricted fashion than is the case for the original data. Since the synthetic data resembles the original data in structure and general characteristics, analysis conducted with it can be useful and meaningful. Results from analysis of synthetic data can be compared with identical analysis conducted on the original data set ("verification service," **PURPLE** hexagon). The service has access to both the highly protected original data and is accessible by individuals with access only to the synthetic data.

Such tailoring and easy adaptation to requirements are made simpler by the virtualization scheme that underlies provisioning resources in the Protected Network. For example, in summer 2017, an instance of Apache Spark (http://spark.apache.org/) was positioned for the DIBBs project in order to pilot its use for fast lookups and distributed processing of large scale data. During earlier stages of the project, such SQL databasing and statistical analysis was handled by a stand-alone server in the Protected Network, and scaling the database resource to meet greater demand and more complex analysis was challenging. The applicability of Apache Spark to the project was not foreseen at the onset of the project, and flexibility of virtualized resources in the Protected Network allowed a "mid-course correction" to take advantage of the emerging technology. With virtualization, new and developing technologies like Spark can be provisioned into the Protected Network very economically, so that the technology can be easily scaled up and tested.

The infrastructure that the DIBBs project has developed uses the Protected Network environment as a protected space to design and implement an innovative research computing environment where data can be exposed to researchers in a manner that is less or more restrictive according to the sensitivity of the data sets and the privileges granted to the researcher. The research group has created "synthetic data" from the original and very sensitive OPM data, and the synthetic data can more readily be made available to researchers, who can get meaningful information from it without disclosing personally identifiable information of federal government employees (Barrientos, *et al*. 2017). Moreover, the underlying IT infrastructure makes verification of results derived from synthetic data possible as well, since results from the synthetic data analysis can be compared with results from the "real" data – done in a manner that is controlled and that records and manages differential privacy measures.

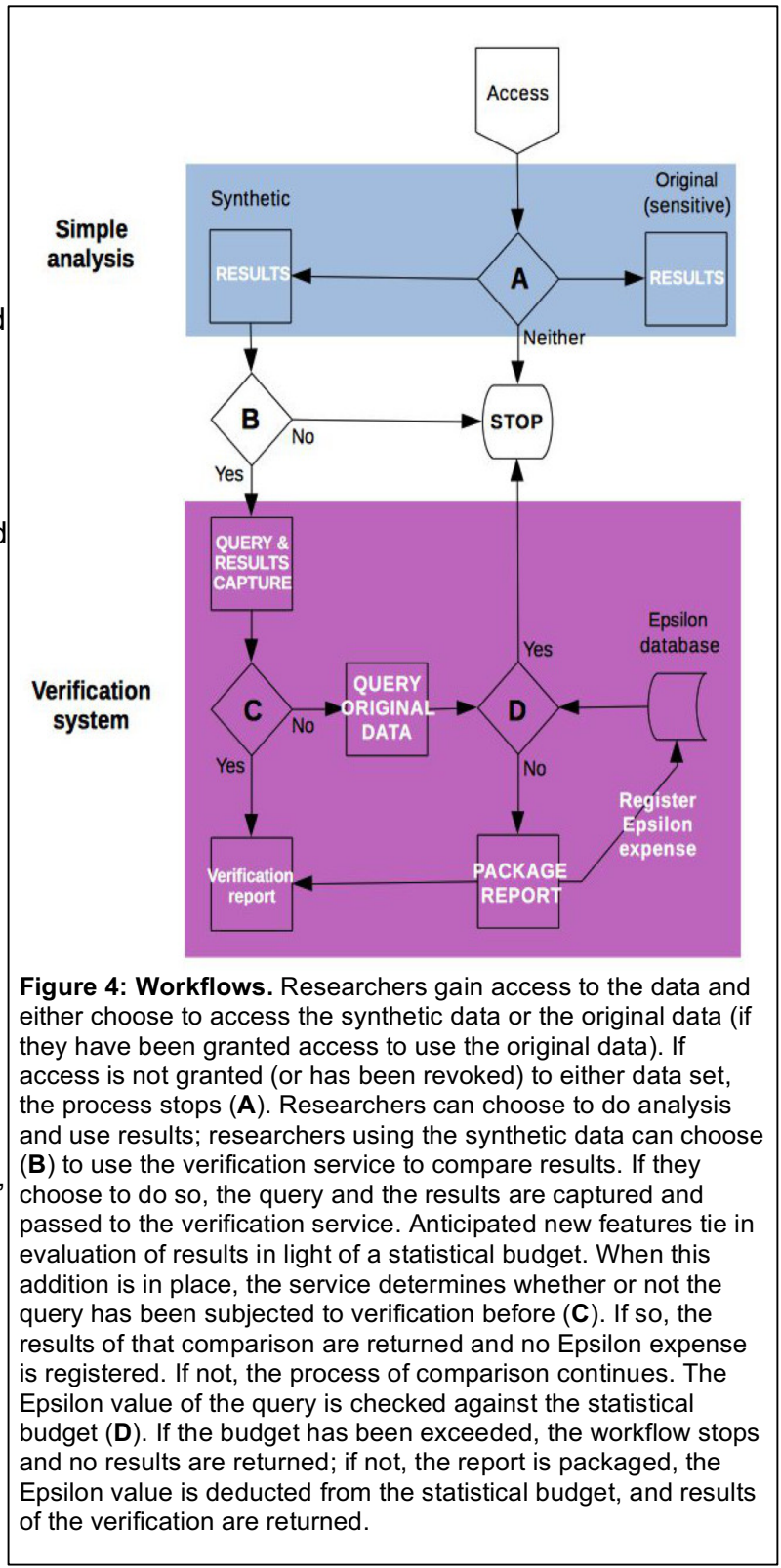

**Figure 4: Workflows.** Researchers gain access to the data and either choose to access the synthetic data or the original data (if they have been granted access to use the original data). If access is not granted (or has been revoked) to either data set, the process stops (**A**). Researchers can choose to do analysis and use results; researchers using the synthetic data can choose (**B**) to use the verification service to compare results. If they choose to do so, the query and the results are captured and passed to the verification service. Anticipated new features tie in evaluation of results in light of a statistical budget. When this addition is in place, the service determines whether or not the query has been subjected to verification before (**C**). If so, the results of that comparison are returned and no Epsilon expense is registered. If not, the process of comparison continues. The Epsilon value of the query is checked against the statistical budget (**D**). If the budget has been exceeded, the workflow stops and no results are returned; if not, the report is packaged, the Epsilon value is deducted from the statistical budget, and results of the verification are returned.

Figure 3 schematically represents the way that computational and data resources are isolated and linked via a "verification server" that compares results from analysis executed on the synthetic (less sensitive) and the "real" (more sensitive) data (Reiter, Oganian, & Karr 2009). Comparisons are relayed back to the researcher without revealing original sensitive data, though the system also takes into account differential privacy measures and can abide by an "analysis budget" to limit information about the original data that may be leaked by computed comparisons of real and synthetic data. Figure 4 illustrates typical workflows that researchers use in the integrated system, including planned, but not yet implemented, provisions for enforcement of a statistical budget.

The entire integrated system allows research conducted on synthetic data to be confidently considered valid of the original data as well, meaning that researchers can make conclusions about the original sensitive data – in this case OPM data – without requiring direct access to that very sensitive data. In practical terms, the system will encourage explorations of less sensitive data, leading to more targeted and efficient interrogation of the original data with less risk of disclosure.

*4.3. Special application: Outfitting a protected environment for medical research*

Protected Health Information originating from the Duke Health System is an important resource for biomedical and institutional research, but it is also highly sensitive and subject to regulations from HIPAA and HITECH. In order maintain compliance with the regulations, minimize risk to the health system, and exploit research opportunities that the data afford, Duke Health and University IT organizations have collaborated to develop a Duke Health System-based environment called PACE ("Protected Analysis Computing Environment"). In many ways, PACE shows a kinship with the Protected Network, though its implementation uses technologies that are more familiar to Duke Health Technology Solutions (DHTS) IT staff.

Virtual machines with applications that are used for most clinical studies are "standard issue" in the environment, and because they are standardized, creation and maintenance is simplified. A greater challenge are resources that are needed for other complex and often "big data" projects that may require GPU processors, a large number of CPU-cores or large memory, or custom (and often experimental) software. Provisioning of these "exceptions" has led to an innovative automation design that includes containerization technology (Docker and Singularity), staged provisioning, and relocation of approved containers in the secure enclave. The systems comprise a tightly controlled variant of a typical "continuous delivery pipeline" that accommodates constraints of a protected network environment.

Containers of machines are developed and vetted outside of PACE, where they are approved by researchers as usable for their sensitive project. Introduction of the approved containerized machine is done in a tightly controlled manner, involving an approval for inclusion into PACE and subsequent automated deployment into the environment. Because the containers comprise machine images, access to external software sources such as the Comprehensive R Archive Network (CRAN; [https://cran.r-project.org/](https://cran.r-project.org/)) from within PACE is not required. Updating of machines within the environment essentially means updating externally and then using the continuous delivery pipeline to deploy inside of PACE.

All containerized software in the PACE environment interacts only with data resources provided by Duke Health-maintained storage devices, and research using sensitive data is conducted with the virtual machines only within PACE. Research results can be removed from PACE through an "honest broker" – an administrative process that applies regulations and data use agreements to the specific data that researchers wish to move outside the environment.

Although PACE is separated (unlike the PRDN) from the Protected Network, the technologies and principles that are in use and in effect in the Protected Network have been reapplied. Indeed, PACE, the Protected Network, and the PRDN are very complementary. Prototypes of the continuous integration pipeline that is essential to deployment of PACE VMs were developed and tested in the Protected Network and PRDN. As the PACE environment develops, we expect that lessons learned will inform the further development of the Protected Network.